\documentclass[12pt]{article}

\usepackage{a4}
\usepackage{pslatex}
\usepackage{hyperref}
\usepackage[latin1]{inputenc}
\usepackage[T1]{fontenc}
\usepackage{graphicx}

\makeatletter
\def\@sect#1#2#3#4#5#6[#7]#8{\ifnum #2>\c@secnumdepth
  \def\@svsec{}\else 
  \refstepcounter{#1}\edef\@svsec{\csname the#1\endcsname.\hskip0.5em}\fi
  \@tempskipa #5\relax
  \ifdim \@tempskipa>\z@
  \begingroup 
     #6\relax
     \@hangfrom{\hskip #3\relax\@svsec}{\interlinepenalty \@M #8\par}%
  \endgroup
  \csname #1mark\endcsname{#7}\addcontentsline
      {toc}{#1}{\ifnum #2>\c@secnumdepth \else
        \protect\numberline{\csname the#1\endcsname}\fi #7}%
  \else
    \def\@svsechd{#6\hskip #3\@svsec #8\csname #1mark\endcsname
      {#7}\addcontentsline{toc}{#1}{\ifnum #2>\c@secnumdepth \else
        \protect\numberline{\csname the#1\endcsname}\fi #7}}%
  \fi \@xsect{#5}}
\@addtoreset{equation}{section}
\@addtoreset{figure}{section}
\makeatother

\renewcommand\thesection{\Roman{section}}

\renewcommand\theequation{%
  \ifnum \value{section}>0
     \thesection.\arabic{equation}%
  \else
     \arabic{equation}%
  \fi}
\renewcommand\thefigure{%
  \ifnum \value{section}>0
     \thesection.\arabic{figure}%
  \else
     \arabic{figure}%
  \fi}

\def\L{\left(}
\def\R{\right)}
\def\LB{\left[}
\def\RB{\right]}
\def\Re{{\rm Re}}
\def\sigBorn{{\sigma_0}}
\def\Afin{{\overline{A}}}
\def\Bfin{{\overline{B}}}
\def\PP#1{\LB{#1}\RB_+}
\def\Li2#1{\mbox{Li}_2\left(#1\right)}
\def\GeV{\mbox{ GeV}}
\def\qb{{\bar q}}
\def\tb{{\bar t}}
\def\kqb{k_{\qb}}
\def\ktb{k_{\tb}}
\def\kg{k_{g}}
\def\sqt{{s_{qt}}}
\def\sqtb{{s_{q\tb}}}
\def\kq{k_{q}}
\def\kt{k_{t}}
\def\m#1{{m_{#1}}}
\def\mt{{m_t}}
\def\mb{{m_b}}
\def\mz{{m_Z}}
\def\mw{{m_W}}
\def\mh{{m_H}}
\def\sw{{s_W}}
\def\cw{{c_W}}
\def\thetaw{{\vartheta_W}}
\def\gw{{g_W}}
\def\gvq{{g_v^q}}
\def\gaq{{g_a^q}}
\def\gvt{{g_v^t}}
\def\gat{{g_a^t}}

\def\CF{{C_F}}

\def\calV{{\cal V}}
\def\as{{\alpha_s}}
\def\e{\varepsilon}
\def\bra{\left\langle}
\def\ket{\right\rangle}
\def\order#1{{\cal O}(#1)}

\def\DEWsix#1{{D^{d=6, EW #1}_{0}}}
\def\Dsix#1{{D^{d=6,#1}_{0}}}
\def\QCDC01234{{C_0^{\rm QCD1}(2,3,4)}}
\def\QCDC02234{{C_0^{\rm QCD2}(2,3,4)}}
\def\QCDC01134{{C_0^{\rm QCD1}(1,3,4)}}
\def\C01124{{C_0^{\rm 1}(1,2,4)}}
\def\C02124{{C_0^{\rm 2}(1,2,4)}}
\def\B0613{{B_0^{6}(1,3)}}
\def\B0114{{B_0^{1}(1,4)}}
\def\B0124{{B_0^{1}(2,4)}}
\def\B0224{{B_0^{2}(2,4)}}

\def\Ioperator{{\bf I}}
\def\Koperator{{\bf K}}
\def\Poperator{{\bf P}}
\def\muq{{\mu^2}}
\def\nn{\nonumber}
\def\Ref#1{Ref.~\cite{#1}}
\def\Refs#1{Refs.~\cite{#1}}
\def\Eq#1{{Eq.~(\ref{#1})}}
\def\Eqs#1{{Eqs.~(\ref{#1})}}

\def\rz{\rho_z}
\def\rh{\rho_H}
\def\rw{\rho_w}
\def\rb{\rho_b}
\def\ys{y_s}
\def\yphi{y_{\phi}}

\begin{document}
\thispagestyle{empty}
\begin{flushright}
  CERN-PH-TH/2005-145\\
  TTP05-12\\
  SFB/CPP-05-36
\end{flushright}
\vspace*{3cm}
\begin{center}
  {\Large\bf
    Electroweak corrections to top-quark pair production in quark--antiquark
    annihilation
    }\\
  \vspace*{1cm}

  J.H.~Kühn$^a$, A. Scharf$^a$, and P. Uwer$^b$\\
  \vspace*{0.5cm}
  {\em $^a$Institut für Theoretische Teilchenphysik,
    Universität Karlsruhe\\ 76128 Karlsruhe, Germany}\\
  {\em $^b$CERN, Department of Physics, Theory Division,\\
    CH-1211 Geneva 23, 
    Switzerland}    
\end{center}
\vspace*{1.5cm}
\centerline{\bf Abstract}
\begin{center}
  \parbox{0.8\textwidth}{
    Top-quark physics plays an important r\^ole at hadron colliders
    such as the Tevatron at Fermilab or the LHC at CERN. Given the planned
    precision at these colliders, precise theoretical predictions are 
    required. In this paper we present the complete electroweak
    corrections to QCD-induced top-quark pair production in 
    quark--antiquark annihilation. In particular we provide compact
    analytic expressions for the differential partonic cross section, 
    which will be useful for further theoretical investigations.
}
\end{center}

\newpage
\setcounter{page}{1}
\section{Introduction}
\label{sec:intro}
At ongoing and upcoming collider experiments, top-quark physics will
play a central r\^ole. Although the top-quark was discovered already 
10 years ago, direct measurements of its properties are still rather 
limited. In particular most of the quantum numbers are only
constrained from indirect measurements such as  
the electroweak precision observables. 
In the near future the hadron colliders Tevatron
at Fermilab and LHC at CERN will provide unique possibilities for
detailed measurements in the top sector. A necessary requirement for 
these analyses is the precise theoretical
understanding of reactions involving top-quarks.  
At hadron colliders both single top-quark production as well as 
top-quark pair production have been studied extensively in the past. The
differential cross section for top-quark pair production is known to 
next-to-leading order (NLO) accuracy in quantum chromodynamics (QCD) 
\cite{Nason:1988xz,Nason:1989zy,Beenakker:1989bq,Beenakker:1991ma,Bernreuther:2001rq}.
In addition, the resummation of logarithmic enhanced contributions has
been studied in detail in 
\Refs{Laenen:1992af,Kidonakis:1995wz,Berger:1996ad,Catani:1996yz,Berger:1998gz,Cacciari:2003fi}.
Recently also the spin correlations between top-quark and antitop-quark
were calculated at NLO in QCD \cite{Bernreuther:2001rq,Bernreuther:2004jv}.
In \Refs{Beenakker:1993yr,Kao:1999kj} the electroweak corrections were
investigated. However, for the quark--antiquark annihilation process, only the 
electroweak vertex corrections were considered --- the contributions
from box diagrams were ignored. It is well known that 
in the high energy region $s\gg m_{W,Z}$  
the weak corrections can be enhanced by the presence of large 
logarithms (see e.g. \Refs{Kuhn:1999nn,Kuhn:2001hz} and references therein)
which justifies a detailed study of all contributions. 
More recently the electroweak corrections for $b$-quark production 
were re-analysed
in \Refs{Maina:2003is,Maina:2004dm}. In particular,
it was again confirmed that the weak corrections can lead to sizable
corrections for specific observables. For a more detailed theoretical
investigation of these effects, it is useful to have short analytic
expressions available. The aim of this work is to recalculate the
weak corrections to top-quark pair production --- including
the contribtution from box diagrams --- and give compact
analytic results. Note that we do not consider here the pure photonic 
corrections, which form a separate gauge-invariant subset.

The outline of the paper is as follows. In section 2 we present
the calculation of the virtual electroweak corrections to top-quark pair
production in quark--antiquark annihilation. The contributions
involving box diagrams are infrared-divergent. The singularities
cancel when the virtual corrections are combined with the
corresponding real corrections, which we calculate in section 3. In 
section 4 we discuss some checks we performed and give numerical
results for the total cross section.
 
\section{Virtual corrections}
\label{sec:virtual}
\begin{figure}[!htbp]
  \begin{center}
    \leavevmode
    \includegraphics[width=14cm]{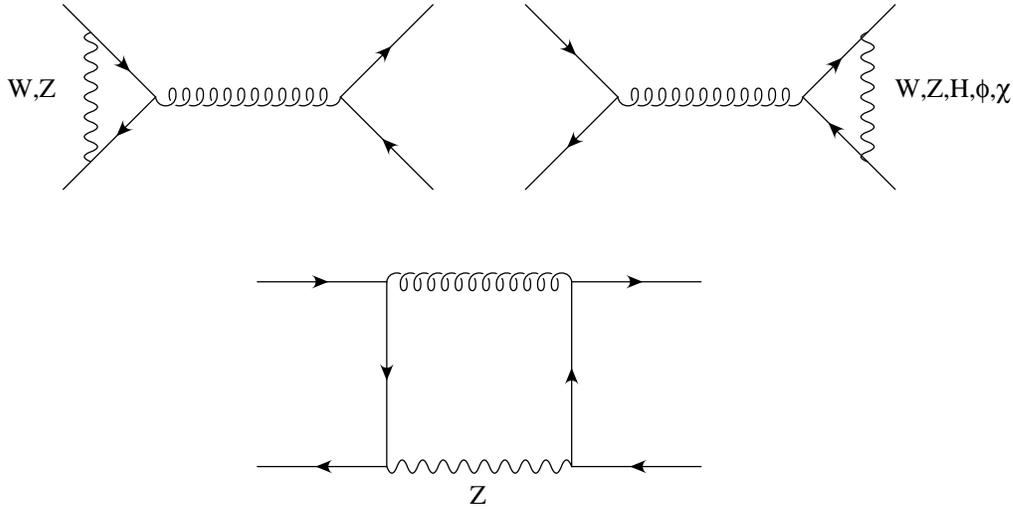}
    \caption{Sample diagrams for the virtual corrections.}
    \label{fig:loop-diagrams}
  \end{center}
\end{figure}

\begin{figure}[!htbp]
  \begin{center}
    \leavevmode
    \includegraphics[width=12cm]{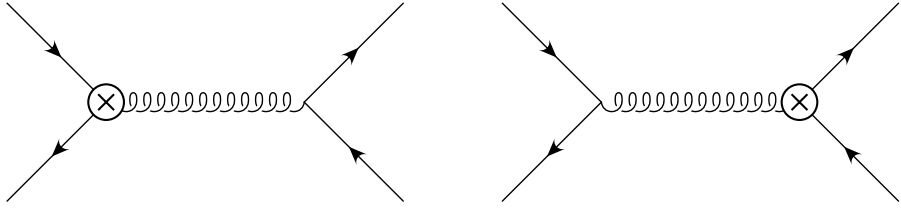}
    \caption{Counterterm diagrams.}
    \label{fig:counterterms}
  \end{center}
\end{figure}

In this section we present the calculation of the electroweak
corrections. We work in the 't Hooft gauge ($R_\xi$-gauge) with the 
gauge parameters $\xi^i$ set to 1. 
In this gauge, apart from the physical fields, also
unphysical fields contribute. 
In particular we have to consider the
contribution from the fields denoted by $\chi$, $\phi$, which are
related to the longitudinal degrees of freedom of the gauge bosons.  
In principle in the $R_\xi$-gauge also ghosts need to be
considered to cancel unphysical degrees  of freedom. To the order
where we
are working, the ghosts do not contribute. In addition, given that we
neglect the  masses of the $u, d, c, s$ quarks the unphysical fields
$\phi$ and $\chi$ only contribute in the vertex corrections to the
final gluon--top--antitop vertex.  The renormalization is done in renormalized
perturbation theory. That is the bare Lagrangian $\cal L$ is rewritten in terms
of renormalized fields and couplings:
\begin{eqnarray}
 {\cal L}(\Psi_0,A_0, m_0,g_0) &=& 
 {\cal L}(Z^{1/2}_\Psi\Psi_R,Z^{1/2}_A A_R, Z_m m_R,Z_g g_R)\nn\\
 &=&  {\cal L}(\Psi_R,A_R, m_R,g_R) + {\cal L}_{ct}(\Psi_R,A_R, m_R,g_R).  
 \label{eq:RenormalizedPerturbationTheory}
\end{eqnarray}
The contribution ${\cal L}(\Psi_R,A_R, m_R,g_R)$ gives just the
ordinary Feynman rules, but with the bare couplings replaced by the
renormalized ones. Some sample diagrams are shown in 
Fig.~\ref{fig:loop-diagrams}. 
The complete list of Feynman rules can be found
for example
in \Ref{Denner:1991kt}. The second contribution   in 
\Eq{eq:RenormalizedPerturbationTheory} ${\cal
  L}_{ct}(\Psi_R,A_R, m_R,g_R)$
yields the counter\-terms,
which render the calculation ultraviolet (UV)-finite. The
diagrams needed here are shown in Fig.~\ref{fig:counterterms}. 
Note that although
the electroweak corrections appear here in one-loop approximation, they
are the  leading-order electroweak contribution. 
The 
interference term of the amplitude ${\cal M}(q\bar q\to \gamma,Z \to
t\bar t)$ with
the corresponding QCD amplitude vanishes as a consequence 
of the specific colour structure. 
Terms of order $\alpha_s \alpha$ are therefore absent.
Thus no renormalization of the coupling constants is required at the order
under consideration here. 
This is different from an electroweak correction to an electroweak
amplitude, 
which would not
be UV-finite without coupling-constant renormalization.
The whole contribution from the renormalization is
given by:
\begin{equation}
  \delta |{\cal M}|^2 = 2 (\delta Z_q + \delta Z_t) |{\cal M}_{q\bar q\to
  t\bar t}|^2,
\end{equation}
where $Z_q$, $Z_t$ denote the wave-function renormalization constants of
the incoming light quark and the outgoing top-quark ($Z_i = 1+\delta Z_i$).   
The squared leading-order QCD amplitude   $|{\cal M}_{q\bar q\to
  t\bar t}|^2$ in $d$ dimensions is given by:
\begin{equation}
  |{\cal M}_{q\bar q\to t\bar t}|^2 = 16\pi^2 \as^2 
  (N^2 - 1) ( 2 - \beta^2(1-z^2) - 2 \varepsilon),  
\end{equation}
where $N$ is the number of colours, $\as$ the strong coupling constant
and $\beta$ the velocity of the top-quark in the partonic
centre-of-mass system:
\begin{equation}
  \beta = \sqrt{1-4{\mt^2\over s}}
\end{equation}
($s$ denotes the partonic centre-of-mass energy squared).
The cosine of the scattering angle is denoted by $z$. The parameter
of dimensional regularization $\e$ is defined by
\begin{equation}
  d = 4 - 2\e.
\end{equation}
For the renormalization of the quark fields we use the on-shell
scheme. The renormalization constants in this scheme in terms of
self-energy integrals and derivatives thereof can be found for example
in \Ref{Denner:1991kt}.
Before presenting the results, let us add a few technical remarks. We
used the Passarino\,--Veltman reduction scheme \cite{PaVe79} to reduce the
tensor integrals to scalar one-loop integrals. 
For the scalar integrals we use the following convention:
\begin{equation}
  X_0 
  = {1\over i\pi^2} \int d^d\ell 
  {(2\pi\mu)^{2\e}\over (\ell^2-m_1^2+i\epsilon)\cdots}.
\end{equation}
For the UV-divergent integrals we define the finite part for the
one-point integrals $A_0$ and the two-point integrals $B_0$ through
\begin{eqnarray}
  A_0(m^2) &=& m^2 \Delta + \overline{A}_0(m^2),\nn\\
  B_0(p^2,m_1^2,m_2^2)    &=&  \Delta+ \overline{B}_0(p^2,m_1^2,m_2^2), 
\end{eqnarray}
with 
\begin{equation}
  \Delta = (4\*\pi)^{\e}\*\Gamma(1+\e)\*{1\over \e}
  = {1\over\e}- \gamma + \ln(4\pi) + O(\e) .
\end{equation}
The vertex corrections do not contain infrared or mass
singularities (IR singularities). 
They contain only UV singularities, which are removed by the
aforementioned renormalization. On the other hand the contribution involving
box diagrams are UV-finite but contain IR singularities.
In order to regularize the IR singularities, we use dimensional regularization.
To simplify their determination, we express
the $d$-dimensional four-point scalar integrals  $D_0^{d}$ in terms of the 
$(d+2)$-dimensional four-point integrals $D_0^{d+2}$ and 
a combination of three-point
integrals in $d$ dimensions. 
This can be done by the following relation
\begin{equation}
  D_0^{d+2} = - 2\pi D^d_{27}
\end{equation}
where the box integral in 6 dimensions is defined by
\begin{eqnarray}
  &&D^{d=6}(p_1^2,p_2^2,p_3^2,p_1\cdot p_2, p_2\cdot p_3,p_1\cdot p_3,
  \m1^2,\m2^2,\m3^2,\m4^2)[1,\ell_\mu,\ell_\mu\ell_\nu,\ldots]  =\nn\\
  &&{1\over i\pi^2} \int d^6\ell \bigg\{
  {[1,\ell_\mu,\ell_\mu\ell_\nu,\ldots]\over (\ell^2-\m1^2+i\e)
    ((\ell+p_1)^2-\m2^2+i\e)}\nn\\
  &&
  \times {1\over ((\ell+p_1+p_2)^2-\m3^2+i\e)
    ((\ell+p_1+p_2+p_3)^2-\m4^2+i\e) }\bigg\}
\end{eqnarray}
and $D^d_{27}$ is the coefficient of the metric tensor $g_{\mu\nu}$ 
appearing in the 
Passarino\,--Veltman decomposition \cite{PaVe79} of the tensor integral
\begin{equation}
  D^{d=6}[\ell_\mu\ell_\nu]
\end{equation}
(see Eq.~(F.3) in \Ref{PaVe79}), 
which in turn can be expressed as a linear combination of the scalar
box integral $D^d_0$ and scalar triangle integrals $C^d_0$ in $d$ dimensions.
Owing to the finiteness of the box
integrals in 6 dimensions the infrared singularities appear only 
in the three-point integrals.  
 
For the presentation of the results it is convenient to use the
leading-order QCD cross section which is given by: 
\begin{equation}
  {d\sigma^{\rm Born}\over dz} =\sigBorn (2-\beta^2+\beta^2 z^2) 
\end{equation}
with
\begin{equation}
  \sigBorn = {1\over 8}\pi \alpha_s^2 {N^2-1\over N^2} {\beta\over s}.
\end{equation}
A factor $1/(4N^2)$ from averaging over the incoming spins and colour
is included. In a first step we present the corrections to the cross
section from the contribution of $Z$ boson and $W$ boson exchange to
the light quark--gluon--vertex. The form of this vertex remains
unchanged, its normalization is shifted by a factor $(1+\delta
F_D^{\rm In, W+Z})$, which leads to a shift by
\begin{equation}
  {d\sigma^{\rm In., W+Z}\over dz} = {d\sigma^{\rm Born}\over dz}
  2 \Re \delta F_D^{\rm In, W+Z}
  \label{eq:InVertices}
\end{equation}
with 
\begin{equation}
   \delta F_D^{\rm In, W+Z} = -{1\over 8}{\alpha\over \pi}\bigg(
  (\gvq^2+\gaq^2) \* f_1(\rz) 
  +  2 \gw^2\* f_1(\rw)\bigg)
  \label{eq:InVertices2}
\end{equation}
and 
\begin{equation}
  f_1(x) = 1+2\*\Big[\big(1+\ln(x)\big)
  \*\big(2\*x+3\big)-2\*\big(1+x\big)^2
  \Big(\Li2{1+{1\over x}}-{\pi^2\over 6}\Big)\Big],
\end{equation}
where we used the definition
\begin{equation}
  \rho_i = {m_i^2\over s}.
\end{equation}
In \Eq{eq:InVertices2}, $\alpha= {e^2\over 4\pi}$ denotes the 
electromagnetic coupling.\\ The Cabibbo\,--Kobayashi--Maskawa
mixing matrix has been set to 1.
The vector and axial vector couplings of neutral and charged currents are
given by
\begin{eqnarray}
  \gvq &=& {1\over 2\sw\cw}( T_3^q - 2 \sw^2 Q_q ),\\
  \gaq &=& {1\over 2 \sw \cw } T_3^q, \\
  \gw  &=& {1\over 2\sqrt{2}  \sw },
\end{eqnarray}
where $Q_q$ describes the electric charge in units of the elementary
charge $e$, $\sw$ is the sine of the Weinberg angle
($\sw = \sin(\thetaw)$ , $\cw = \cos(\thetaw)$), and $T_3^q$ denotes the weak isospin.
For the vertex corrections to the final vertex we split the result
into the contribution from $Z$ boson exchange, $W$ boson exchange, 
Higgs exchange and the contributions from the would-be Goldstone
bosons including the respective counter terms:
\begin{equation}
  d\sigma^{\rm Fin.} = d\sigma^{{\rm Fin.}, Z} 
  + d\sigma^{{\rm Fin.}, W} 
  + d\sigma^{{\rm Fin.}, H} 
  + d\sigma^{{\rm Fin.}, \chi} 
  + d\sigma^{{\rm Fin.}, \phi}. 
\end{equation}
Note that only the sum has a physically meaningful interpretation.
For the individual contributions we obtain
\begin{eqnarray}
  &&{d\sigma^{{\rm Fin.}, Z}\over dz} = 
  \sigBorn\*{\alpha\over 4\pi}
  \* \Bigg\{ - 2\*(\gvt^2+\gat^2)\*(1+z^2) \nn \\ 
  &+& 
  8 \* {1+z^2\over (1-\beta^2)\*s}\*(\gvt^2+\gat^2)\*
  (\Afin_0(\mz^2)-\Afin_0(\mt^2)) \nn \\ 
  &-& 4 \* \Bigg[{1\over (1-\beta^2)\*\beta^2 }
  \* \Big(1-\beta^2+2\*\beta^4\*(1-z^2)+7\*\beta^2\*z^2-3\*z^2\Big)\*\rz\*
  (\gat^2+\gvt^2) \nn \\
  && + \Big(1-5\*z^2-3\*\beta^2\*(1-z^2)\Big)
  \*\gat^2
  - \Big(3+z^2-\beta^2 \* (1-z^2) \Big)\*\gvt^2 
  \Bigg]\*\Bfin_0^1(3,4) \nn \\
  &+& 2\*
  \Bigg[{2\over \beta^2}\*f_2(z,\beta)\*\rz\*(\gat^2+\gvt^2)
   - f_2(z,\beta)\* (\gvt^2 -3\*\gat^2) 
  - 4\*(1+z^2)\*\gvt^2\Bigg]\*\Bfin_0^3(1,3) \nn \\
  &+& 2\*s\*
  \Bigg[4\*\Big(f_2(z,\beta)\*\gat^2
  -\Big(1+z^2\Big)\*\gvt^2\Big)\*\rz + {2 \over \beta^2}\*f_2(z,\beta)
  \*\rz^2\*(\gvt^2+\gat^2) \nn \\
  &-&(1+\beta^2)\*(2-\beta^2\*(1-z^2))\*\gvt^2-
  \Big(2-3\*\beta^2+5\*\beta^2\*z^2+3\*\beta^4\*(1-z^2)\Big)\*\gat^2
 \Bigg]\*C_0^3 \nn\\ 
  &+& 2\*s\*
  \Big[2\*\rz\*(\gvt^2+\gat^2)
  +(1-\beta^2)\*\L \gvt^2
  -3\*\gat^2\R  
  \Big]\nn\\
  &\times&\*\Big(2-\beta^2\*(1-z^2)\Big) \* {d\over dp^2}
  \left.B_0^1(3,4)\right|_{p^2=\mt^2}\Bigg\},
\end{eqnarray}
with 
\begin{equation}
  f_2(z,\beta) = 1-3\*z^2-2\*\beta^2\*(1-z^2).
\end{equation}
The integrals are defined in the appendix. 
For the contribution from the $W$ boson we obtain:
\begin{eqnarray}
  && {d\sigma^{{\rm Fin.}, W}\over dz} = \sigBorn {\alpha\over 2\pi}\*\gw^2
  \Bigg\{
  -2\*(1+z^2)
  + 8 \* {1+z^2\over (1-\beta^2)\*s} \* (\Afin_0(\mw^2)-\Afin_0(\mb^2))\nn \\
  &-& {1\over \beta^2}\*
  \Big[{4\over (1-\beta^2)}
  \*\Big(1-3\*z^2-\beta^2+7\*\beta^2\*z^2 + 2\*\beta^4\*(1-z^2)\Big)
  \*(\rw-\rb) \nn \\
  &&
  +1-3\*z^2-5\*\beta^2\*(1+z^2)-2\*\beta^4\*(1-z^2)\Big]
  \*\Bfin_0^4(1,2) \nn \\
  &+& {1\over \beta^2}\*\Big[4\*f_2(z,\beta)
  \*(\rw-\rb)
  + 1-3\*z^2-5\*\beta^2\*(1+z^2)-2\*\beta^4\*(1-z^2)\Big]\*\Bfin_0^4(1,3)
  \nn \\
  &+& {s\over 4\*\beta^2}\*
  \Big[8\*f_2(z,\beta)\*(2\*(\rw-\rb)^2 -(1+\beta^2)\*\rb)
  +1-3\*z^2-16\*\beta^2-4\*\beta^2\*z^2+\beta^4\nn\\
  &-& 
  11\*\beta^4\*z^2
  -2\*\beta^6\*(1-z^2)
  +8\*\Big(1-3\*z^2-2\*\beta^4\*(1-z^2)
  -3\*\beta^2\*(1+z^2)\Big)\* \rw \Big]\*C_0^4 \nn \\
  &+&s\*\Big(2-\beta^2\*(1-z^2)\Big)\*\Big[4\*(\rw-\rb)-(1-\beta^2)\Big]
  \*{d\over dp^2} \left.B_0^4(1,2)\right|_{p^2=\mt^2}\Bigg\}.
\end{eqnarray}
The contribution from the Higgs boson is given by
\begin{eqnarray}
  && {d\sigma^{{\rm Fin.}, H}\over dz} =\sigBorn {\alpha\over 2\pi} 
  \*\gw^2\*{\mt^2\over\mw^2}\*\Bigg\{
  -(1+z^2)\nn\\ 
  &+&4\*{1+z^2\over (1-\beta^2)\*s}\*(\Afin_0(\mh^2)-\Afin_0(\mt^2)) \nn \\
  &-&2
  \*\Big[ {1\over(1-\beta^2)\*\beta^2}\*
  \Big(1-\beta^2+2\*\beta^4\*(1-z^2)+7\*\beta^2\*z^2-3\*z^2\Big)
  \*\rh \nn\\
  &+&2\*(1-z^2)\*(1-\beta^2)\Big]\*\Bfin_0^5(1,2) \nn\\
  &+&{1\over\beta^2}\*\Big[2\*f_2(z,\beta)\*\rh
  +\beta^2\*\Big(5-3\*z^2-4\*\beta^2\*(1-z^2)\Big)\Big]\*\Bfin_0^3(1,3)\nn\\
  &+&{2\*s\over\beta^2}\*\Big[f_2(z,\beta)\*\rh^2
  +3\*\beta^2\*(1-z^2)\*(1-\beta^2)\*\rh
  -\beta^2\*(1-\beta^2)\*\Big(2-\beta^2\*(1-z^2)\Big)\Big]\*C_0^5\nn\\
  &+&2\*s\*\Big(2-\beta^2\*(1-z^2)\Big)\*\Big[\rh-(1-\beta^2)\Big]
  \*{d\over dp^2 } \left.B_0^5(1,2)\right|_{p^2=\mt^2}\Bigg\}.
\end{eqnarray}
For the unphysical fields $\chi$ and $\phi$ we find
\begin{eqnarray}
  &&{d\sigma^{\rm Fin., \chi}\over dz} = \sigBorn {\alpha\over 2\pi}
     \*\gat^2
  {\mt^2\over\mz^2}
  \*\Bigg\{-2\*(1+z^2) 
  -
  8\*{1+z^2\over (1-\beta^2)\*s}\*(\Afin_0(\mt^2)-\Afin_0(\mz^2))\nn\\
  &+&{2\over \beta^2}\*\Big[2\*f_2(z,\beta)\*\rz
  +\beta^2\*(1+z^2)\Big]\*\Bfin_0^3(1,3) \nn\\
  &-&{4\over\beta^2\*(1-\beta^2)}
  \*\Big[1-\beta^2+2\*\beta^4\*(1-z^2)+7\*\beta^2\*z^2-3\*z^2\Big]
  \*\rz\*\Bfin_0^1(3,4) \nn\\
  &+&{4\*s\over \beta^2}\*\Big[f_2(z,\beta)\*\rz^2
  +\beta^2\*(1-z^2)\*(1-\beta^2)\*\rz\Big]\*C_0^3\nn\\
  &+&4\*s\*\Big(2-\beta^2\*(1-z^2)\Big)\*\rz\*{d\over dp^2}
  \left. B_0^1(3,4)\right|_{p^2=\mt^2} 
  \Bigg\},
\end{eqnarray}
\begin{eqnarray}
  &&d\sigma^{\rm Fin.,\phi} = \sigBorn\*{\alpha\over2\pi}\*\gw^2\Bigg\{
  -{1\over 4\*\rw}\*\ys\*(1+z^2)
  +{(1+z^2)\over(1-\beta^2)\*\mw^2}\*\ys\*(\Afin_0(\mw^2)
  -\Afin_0(\mb^2)) \nn \\
  &-&{1\over8\*\beta^2\*(1-\beta^2)}\*
  \Big[4\*\Big(1-3\*z^2-\beta^2+7\*\beta^2\*z^2+2\*\beta^4\*(1-z^2)\Big)\*\ys 
  \nn \\
  &-&16\*\Big(1-3\*z^2-\beta^2+7\*\beta^2\*z^2+2\*\beta^4\*(1-z^2)\Big)\*\rb
  \*\yphi
  -16\*\beta^2\*(1-\beta^2)^2\*(1-z^2)\*\yphi \nn \\
  &+&(1-\beta^2)^2\*
  \Big(1-3\*z^2+3\*\beta^2\*(1+z^2)-2\*\beta^4\*(1-z^2)\Big)\*\rw^{-1}\Big]
  \*\Bfin_0^4(1,2) \nn \\
  &+&
  {1\over8\*\beta^2}\*\Big[4\*f_2(z,\beta)\*(\ys
  -4\yphi\*\rb -2\*\beta^2\*\yphi)\nn \\
  &+&(1-\beta^2)\*
  \Big(1-3\*z^2+3\*\beta^2\*(1+z^2)-2\*\beta^4\*(1-z^2)\Big)\*\rw^{-1}
  \Big]
  \*\Bfin_0^4(1,3) \nn \\
  &+&{s\over32\*\beta^2}\*
  \bigg[16\*f_2(z,\beta)\*(\ys\*\rw
  +4\*\rb^2\*\yphi-8\*\rb^2)
  +8\*(1-\beta^2)^2\*\Big(1-3\*z^2+2\*\beta^2\*(1-z^2)\Big) \nn \\
  &-&16\*
  \Big(1-3\*z^2+\beta^2+9\*\beta^2\*z^2+2\*\beta^4\*(1-z^2)\Big)
  \*\rb\*\yphi \nn \\
  &-&4\*(1-\beta^2)\*
  \Big(1-3\*z^2-11\*\beta^2-3\*\beta^2\*z^2+2\*\beta^4\*(1-z^2)\Big)
  \*\yphi \nn \\
  &+&
  (1-\beta^2)^2\*\Big(1-3\*z^2-7\*\beta^2+\beta^2\*z^2
  +2\*\beta^4\*(1-z^2)\Big)
  \*\rw^{-1}\bigg]\*C_0^4\nn \\
  &+&{s\over 8}
  \*\Big[4\*\ys-16\*\rb\*\yphi+8\*(1-\beta^2)\*\yphi
  -{(1-\beta^2)^2\over \rw}\Big]\nn\\
  &&\times\Big(2-\beta^2\*(1-z^2)\Big)\*{d\over d p^2} 
  \left.B_0^4(1,2)\right|_{p^2=\mt^2}
  \Bigg\},
\end{eqnarray}
where we used the abbreviations:
\begin{eqnarray}
&&\ys = 1-\beta^2+4\rb, \\
&&\yphi = {\mb^2\over \mw^2}.
\end{eqnarray}
Again only terms proportional to $\gvt^2, \gat^2$ or $\gw^2$ are present.
Let us now discuss the contribution from the box diagrams. To the
order $\alpha \as^2$ considered  here, we can distinguish
two different contributions:
\begin{enumerate}
\item The (box-type) electroweak correction to the QCD Born amplitude, 
  interfering with the QCD Born amplitude.
\item The QCD box diagram interfering with the electroweak Born
  amplitude.
\end{enumerate}
In the following we will call the first the
electroweak-box (EW-box), and the second the QCD-box contribution.
For the EW-box we obtain
\begin{eqnarray}
  &&{d\sigma^{\rm EW-box }\over dz}=
  \sigBorn\*{\alpha\over \pi}\*\Bigg\{
    -{2\*(1-\beta^2)\over \beta\*(1-\beta^2\*z^2)}\*
  \Bigg[
  z\*\bigg(1+2\*\beta^2-\beta^2\*z^2\bigg)\*\gvq\*\gvt
  +2\*\beta\*\gaq\*\gat
  \Bigg]\*\Bfin_0^1(1,4)
  \nn\\
  &+&{2\*(1-2\*\beta^2+\beta^2\*z^2)\over 1-\beta\*z}\*
  \Bigg[\gvq\*\gvt+\gaq\*\gat\bigg]\*\Bfin_0^1(2,4)
  +{4\over \beta}\*\Bigg[\gvq\*\gvt\*z+\gaq\*\gat\*\beta\Bigg]\*\Bfin_0^1(1,3)
  \nn\\
  &-&{2\*(1-\beta^2)\over \beta\*(1-\beta^2\*z^2)}
  \*\Bigg[
  z\*\bigg(1+2\*\beta^2-\beta^2\*z^2\bigg)\*\gvq\*\gvt
  +2\*\gaq\*\gat\*\beta
  \Bigg]
  \*\Bfin_0^1(3,4)
  \nn\\
  &-&{2\over 1+\beta\*z}
  \*(1-2\*\beta^2+\beta^2\*z^2)\*
  \Bigg[\gvq\*\gvt-\gaq\*\gat\Bigg]\*\Bfin_0^2(2,4)
  \nn\\
  &+&s\*{2\*(1-\beta^2)\over 
    \beta \* (1-\beta^2\*z^2)\* (1-\rho_z)}\*
  \Bigg[
  z\*\bigg(1+\beta^4\*z^2-\beta^2\*z^2-\beta^4+2\*\beta^2\nn\\
  &+&(\beta^4\*z^2+2\*\beta^2\*z^2-2-\beta^4)\*\rho_z
  -(-2\*\beta^2-1+\beta^2\*z^2)\*\rho_z^2
  \bigg)\*\gvq\*\gvt\nn\\
  &+&2\*\beta\*\bigg(\beta^2\*z^2
  -(1-\beta^2\*z^2)\*\rho_z+\rho_z^2\bigg)\*\gaq\*\gat
  \Bigg]
  \*C_0(1,3,4)
  \nn\\
  &+& s \* {1\over (1-\beta\*z)\*(1-\rho_z)^2}
  \*\Bigg[
  \bigg(2\*(\beta^2\*z^2-2\*\beta^2+1)\*\rho_z^3\nn\\
  &-&
  (4-5\*\beta^2+2\*\beta^3\*z+\beta^2\*z^2-2\*\beta^3\*z^3
  -\beta^4+\beta^4\*z^2)\*\rho_z^2\nn\\
  &+&
  \beta\*(\beta^3\*z^4+4\*z-2\*\beta^3\*z^2-4\*\beta+\beta^3)\*\rho_z
  +3\*\beta^2\*z^2-2\*\beta^3\*z^3\nn\\
  &+& 2-4\*\beta\*z-\beta^2-\beta^4\*z^2+\beta^4\*z^4
  +2\*\beta^3\*z\bigg)\*\gvq\*\gvt
  \nn\\
  &+&2\*\bigg((\beta^2\*z^2-2\*\beta^2+1)\*\rho_z^3
  +(-3+4\*\beta^2-2\*\beta^3\*z+\beta^3\*z^3-\beta^2\*z^2+\beta\*z)\*\rho_z^2
  \nn\\
  &+&(3-3\*\beta\*z+\beta^3\*z-3\*\beta^2+2\*\beta^2\*z^2)\*\rho_z
  -2\*\beta^2\*z^2+\beta^2-\beta^3\*z+\beta^3\*z^3\nn\\
  &+&2\*\beta\*z-1\bigg)\*\gaq\*\gat
  \bigg]\*C_0^1(2,3,4)
  \nn\\
  &-&s\*
  {1\over (1+\beta\*z)\*(1-\rho_z)^2}
  \*\Bigg[
  \bigg(
  3\*\beta^2\*z^2+2\*\beta^3\*z^3+2+4\*\beta\*z-\beta^2-\beta^4\*z^2\nn\\
  &
  +&\beta^4\*z^4-2\*\beta^3\*z  
  +\beta\*(\beta^3\*z^4-4\*z-2\*\beta^3\*z^2-4\*\beta+\beta^3)\*\rho_z
  \nn\\
  &-&(4-5\*\beta^2-2\*\beta^3\*z+\beta^2\*z^2+2\*\beta^3\*z^3-\beta^4
  +\beta^4\*z^2)\*\rho_z^2
  \nn\\
  &+&2\*(\beta^2\*z^2-2\*\beta^2+1)\*\rho_z^3
  \bigg)\*\gvq\*\gvt
  +2\*\bigg(
  +2\*\beta^2\*z^2-\beta^2-\beta^3\*z+\beta^3\*z^3+2\*\beta\*z+1
  \nn\\
  &+&(-3-3\*\beta\*z+\beta^3\*z+3\*\beta^2-2\*\beta^2\*z^2)\*\rho_z
  \nn\\
  &+&(3-4\*\beta^2-2\*\beta^3\*z+\beta^3\*z^3+\beta^2\*z^2+\beta\*z)\*\rho_z^2
  \nn\\
  &-&(\beta^2\*z^2-2\*\beta^2+1)\*\rho_z^3
  \bigg)\*\gaq\*\gat
\Bigg]
  \*C_0^2(2,3,4)\nn\\
  &+&{2\*\beta^2\*(1-z^2)\over (1-\beta\*z)\*(1-\rho_z)^2}
  \*\Bigg[
  \bigg(3-\beta^2-2\*\beta\*z\*(1-\beta\*z)
  -(1-2\*\beta\*z+\beta^2)\*\rho_z+ 2\*\rho_z^2
  \bigg)\*\gvq\*\gvt
  \nn\\
  &+&2\*\big(\beta\*z-(1-\beta\*z)\*\rho_z+\rho_z^2\big)\*\gaq\*\gat
  \Bigg]
  \*s\*\DEWsix1
  \nn\\
  &-&{2  \*\beta^2\*(1-z^2)\over
    (1+\beta\*z)\*(1-\rho_z)^2}
  \*\Bigg[
  \bigg(3-\beta^2+2\*\beta\*z\*(1+\beta\*z)
  -(1+\beta^2+2\*\beta\*z)\*\rho_z+2\*\rho_z^2\bigg)\*\gvq\*\gvt
  \nn\\
  &+&2\*\bigg(\beta\*z+(1+\beta\*z)\*\rho_z-\rho_z^2\bigg)\*\gaq\*\gat
  \Bigg]
  \*s\*\DEWsix2
  \Bigg\}
  \nn\\
 &-&{1\over 64\*\pi}\*\as\*{1\over N^2} \*\beta\*B\*
 \bigg[(1-\beta\*z)\*C_0^1(1,2,4)-(1+\beta\*z)\*C_0^2(1,2,4)\bigg] 
 + \order{\e},
 \label{eq:EWBOX}
\end{eqnarray}
with 
\begin{equation}
  B =-8\* \pi\*\alpha\*\as\* (N^2-1)\* { s\over s-\mz^2}
  \*\bigg((d-2-\beta^2\*(1-z^2))\*\gvq\*\gvt
  +\beta\*z\*(d-2)\*(d-3)\*\gaq\*\gat\bigg)
.\label{eq:Bdef}
\end{equation}
Note that the contribution $d\sigma^{\rm EW-box }$ is UV-finite, as 
already mentioned. This can be easily checked by replacing the $\Bfin_0$
integrals by $\Delta$ and verifying that this contribution indeed
vanishes.
Since we are using box integrals in $d=6$ dimensions, the
IR singularities appear only in the three-point integrals. In
particular, in the above result, only the last line in \Eq{eq:EWBOX}
is 
singular. As a consequence only this term needs to be evaluated in $d$ 
dimensions. Note that owing to the structure of the IR-singularities in QCD we
will not pick up finite terms of the form $\e/\e$. They will cancel with 
the corresponding terms in the real corrections, as we will show in
the next section, where the real corrections are discussed.  

For the QCD-box we find
\begin{eqnarray}
  {d\sigma^{\rm QCD}\over dz} &=&
  \sigBorn {\alpha\over \pi}
  \*{1\over
    (1-\rz)}
  \Bigg\{
  +{2\over \beta}\*\Bigg[\beta\*\gaq\*\gat+z\*\gvq\*\gvt\Bigg]\*
  \Bfin_0^6(1,3)
  \nn\\
  &+&2\*{1-\beta^2\over \beta\*(1-\beta^2\*z^2)}\*
  \Bigg[z\*\bigg(-1-2\*\beta^2+\beta^2\*z^2\bigg)\*\gvq\*\gvt
  -\beta\*\bigg(1+\beta^2\*z^2\bigg)\*\gaq\*\gat\Bigg]
  \*\Bfin_0^1(1,4)
  \nn\\
  &+&{1\over 1-\beta\*z}\*
  \Bigg[\bigg(1-2\*\beta^2+\beta^2\*z^2\bigg)\*\gvq\*\gvt
  -\beta\*(\beta-z)\*(1+\beta\*z)\*\gaq\*\gat\Bigg]
  \*\Bfin_0^1(2,4)
  \nn\\
  &-&{1\over 1+\beta\*z}\*
  \Bigg[\bigg(1-2\*\beta^2+\beta^2\*z^2\bigg)\*\gvq\*\gvt
  +\beta\*(\beta+z)\*(1-\beta\*z)\*\gaq\*\gat\Bigg]
  \*\Bfin_0^2(2,4)\nn\\
  &+&{1-\beta^2\over \beta\*(1-\beta^2\*z^2)}\*
  \Bigg[
  z\*\bigg(1+\beta^2+\beta^2\*(1-\beta^2)\*(1-z^2)\bigg)\*\gvq\*\gvt\nn\\
  &+&\beta\*\bigg(1+\beta^2\*z^2\bigg)\*\gaq\*\gat
  \Bigg]\*s\*\QCDC01134\nn\\
  &-&{\beta^2\*(1-z^2)\over 1+\beta\*z} \*
  \Bigg[\bigg(3+2\*\beta\*z\*(1+\beta\*z)-\beta^2\bigg)\*\gvq\*\gvt
  +2\*\beta\*z\*\gaq\*\gat\Bigg]
  \*s\*\Dsix2
  \nn\\
  &+&{\beta^2\*(1-z^2)\over 1-\beta\*z}\*
  \Bigg[\bigg(3-2\*\beta\*z\*(1-\beta\* z)-\beta^2\bigg)\*\gvq\*\gvt
  +2\*\beta\*z\*\gaq\*\gat\Bigg]
  \*s\*\Dsix1
  \Bigg\}
  \nn\\
  &-& {1\over 64\*\pi}\*\as\*{1\over N^2}\*\beta\*B\*\bigg[
  (1-\beta\*z)\*C_0^1(1,2,4)
  -(1+\beta\*z)\*C_0^2(1,2,4)\bigg] 
  + \order{\e}.
  \label{eq:QCDBOX}
\end{eqnarray}
Combining the two results in \Eqs{eq:EWBOX} and (\ref{eq:QCDBOX}), the 
IR-divergent part is 
thus given by 
\begin{eqnarray}
  &-& {\as\over 32\*\pi}\*{1\over N^2}\*\beta\*B\*\bigg[
  (1-\beta\*z)\*C_0^1(1,2,4)
  -(1+\beta\*z)\*C_0^2(1,2,4)\bigg] \nn\\
  &=&
  - {\as\over 16\*\pi}\*{1\over N^2}\*{\beta\over s}\*B\*
  \bigg[\sqt\*C_0^1(1,2,4)-\sqtb\*C_0^2(1,2,4)\bigg]
\end{eqnarray}
In particular, we obtain 
\begin{eqnarray}
  &&- {\as\over 16\*\pi}\*{1\over N^2}\*{\beta\over s}\*B\*
  \bigg[\sqt\*C_0^1(1,2,4)-\sqtb\*C_0^2(1,2,4)\bigg]\nn\\
  &=& 
  {\as\over 32\*\pi}\*{1\over N^2}\*{\beta\over s}\*
  \*B\*(4\pi)^\e\Gamma(1+\e)\bigg[
  {2\over \e}\ln\L {\sqtb\over\sqt}\R 
  + g(\sqt)
  - g(\sqtb)
\bigg],
\label{eq:IRdivcontribution}
\end{eqnarray}
with $s_{ij} = 2 k_i\cdot k_j$,  where we have used
\begin{eqnarray}
  &&C_0^1(1,2,4)
  =
  {1\over i\pi^2}\int d^d\ell {1\over \ell^2(\ell+k_q)^2((\ell+k_t)^2-\mt^2)}
  \nn\\
  &&=
    -{1\over 2}(4\pi)^\e\Gamma(1+\e)
  {1\over \sqt}
  \bigg( 
  {1\over \e^2}
  +{1\over \e}\ln\L {\mt^2\mu^2\over \sqt^2}\R 
  + g(\sqt)
  \bigg) + \order{\e},
\end{eqnarray}
with 
\begin{equation}
  g(\sqt) =   -\ln\L {\sqt\over \mu^2}\R \*\ln\L {\mt^2\over \sqt}\R 
  +{1 \over 2} \*\ln\L {\sqt\over \mu^2}\R ^2
  -2\*\Li2{{\sqt - \mt^2\over \sqt }}
  -{1 \over 2} \*\ln\L {\mt^2\over \sqt}\R ^2.
\end{equation}
We note that the divergent three-point integral 
\begin{equation} 
  {1\over i\pi^2} \int {1\over \ell^2(\ell+\kq)^2
    ((\ell+\kq+\kqb)^2-\mz^2)},
\end{equation}
which could in principle also appear, cancels in the calculation. 

\section{Real corrections}
\label{sec:real}
\begin{figure}[!htbp]
  \begin{center}
    \leavevmode
      \includegraphics[width=12cm]{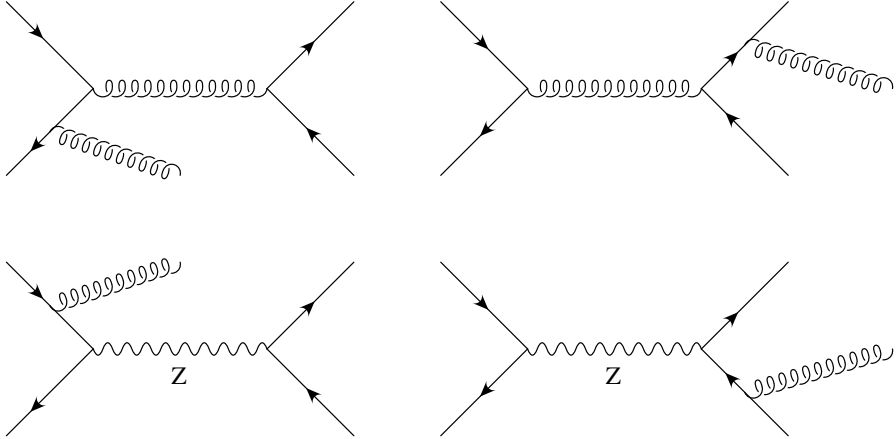}
    \caption{Sample diagrams for the real corrections.}
    \label{fig:RealCor}
  \end{center}
\end{figure}
\begin{figure}[!htbp]
  \begin{center}
    \leavevmode
      \includegraphics[width=5cm]{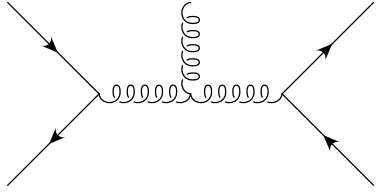}
    \caption{Amplitude containing the triple gluon vertex. The diagram
    does not contribute here because of the colour structure.}
    \label{fig:TripleGluon}
  \end{center}
\end{figure}

As mentioned in the previous section the contribution from the box diagrams
is IR-divergent. To render the corrections to the total cross section finite 
we need to include the real corrections at the same order. A few sample 
diagrams are shown in Fig.~\ref{fig:RealCor}. The diagram containing the
triple gluon vertex (see Fig.~\ref{fig:TripleGluon}) does not contribute because of the colour structure. 
The calculation of the real corrections is straightforward. 
The phase-space integration over the regions where the emitted gluon is soft
will produce the IR singular contribution needed to cancel the corresponding
singularities in the virtual corrections. Note that owing to  
the colour structure
no collinear singularities appear, because the interference between
the two 
diagrams, where the gluon is emitted from the initial state, 
vanishes.
As a consequence no factorization of initial-state singularities is required. 
To extract the IR divergences, we use the so-called subtraction 
method \cite{Catani:1996vz,Phaf:2001gc,Catani:2002hc}.
The basic idea of the subtraction method is to add and subtract a term in such 
a way that the singularities appearing in the real corrections are
matched point-wise and that the term is simple enough to be 
integrated analytically in $d$
dimensions over the full phase space. 
Given that the same term is added and subtracted, this procedure
does not change the result. The analytically integrated term is
combined with the virtual corrections, 
while the unintegrated term is combined with the real corrections.
Given that the term combined with the real corrections match
point-wise the singularities of the squared matrix element, the
integration can be done numerically in 4 dimensions.
Because of the universal structure of soft and mass singularities
in QCD, the subtraction terms can be constructed in a very general way. For
further details on the subtraction method, we refer to 
\Refs{Catani:1996vz,Catani:2002hc}. Here we just
reproduce the necessary equations required for the case at hand.

Using the subtraction method the NLO contribution to the cross section 
can be symbolically written
as \cite{Catani:1996vz,Catani:2002hc}
\begin{eqnarray}
  &&\sigma_{\rm NLO}(\kq,\kqb) = \sigma_{\rm V}(\kq,\kqb)
  + \sigma_{\rm R}(\kq,\kqb) =\nn\\
  &&\hspace{-0.5cm} \int_3 \Bigg\{
  \left[d\sigma_{\rm R}(\kq,\kqb,\kt,\ktb,\kg)\right]_{\e = 0}
  - \left[\sum_{\rm dipoles} 
    \L d\sigma_{\rm LO} \otimes 
    dV_{\rm dipole}\R (\kq,\kqb,\kt,\ktb,\kg)\right]_{\e=0}\Bigg\}\nn\\
  &+& \int_2 \left[ d\sigma_{\rm V}(\kq,\kqb,\kt,\ktb)
  +d\sigma_{\rm LO}(\kq,\kqb,\kt,\ktb) 
  \otimes (\Ioperator_q+\Ioperator_\qb) \right]_{\e = 0}\nn\\
 &+& \int dx \int_2 d\sigma_{\rm LO}(x\kq,\kqb,\kt,\ktb)\otimes (
  \Koperator_q
+ \Poperator_q )\nn\\
 &+& \int dx \int_2 d\sigma_{\rm LO}(\kq,x\kqb,\kt,\ktb)\otimes (
  \Koperator_\qb
+ \Poperator_\qb ).
\label{eq:Master}
\end{eqnarray}
Here $d\sigma_{\rm R}$, $d\sigma_{\rm V}$ denote the real and virtual 
corrections to the cross section. 
In particular we have
\begin{equation}
  d\sigma_{\rm V} = d\sigma^{\rm In.} + d\sigma^{\rm Fin.}
  + d\sigma^{\rm EW-box} + d\sigma^{\rm QCD-box}.
\end{equation}
In \Eq{eq:Master} we label the integral symbols with an index 2 or 3 to
indicate that the phase-space integral runs over a 2- or 3-particle
final state.
The terms of the form
\begin{equation}
  d\sigma_{\rm LO} \otimes {\bf F}
\end{equation}
with ${\bf F} = \Poperator, \Koperator, dV_{\rm dipole}$ deserve some 
explanation. In general the symbol `$\otimes$'
introduces spin as well as colour correlation between the operator $F$
and the leading-order amplitude, which is a vector in colour space. 
Note that for the case studied here, where the gluon is always
emitted from a quark line, no spin correlation appears.
For the contribution from the integrated dipoles we obtain 
\begin{eqnarray}
  &&d\sigma_{\rm LO}(\kq,\kqb,\kt,\ktb) \otimes {\cal I} = 
  {1\over 32\pi} {1\over 4N^2} {\beta\over s}
  \bra q\qb\to t\tb \left| \Ioperator_q+\Ioperator_\qb  \right|
 q\qb\to t\tb \,\ket_{\alpha\as}\nn\\
 &=&- {\as\over 32\pi^2 } {1\over 4N^2} {\beta\over s} 
  {(4\pi)^\e\over \Gamma(1-\e)}\bigg\{
   {2\over \e} \ln\L{\sqtb \over \sqt }\R
   + H(\sqt) - H(\sqtb)
  \bigg\}\nn\\ 
 &\times& 
  \bra q\qb\to t\tb \left| T_q T_t   
  \right| q\qb\to t\tb \,\ket,
  \label{eq:Icontr}
\end{eqnarray}
with
\begin{eqnarray}
    H(\sqt) &=&  -{1 \over 2} \*\ln\L {\mt^2\over \sqt}\R ^2
  -\ln\L {\mt^2\over \sqt}\R\*\ln\L {\sqt\over Q^2(\sqt,\mt)}\R \nn\\
  &-&\ln\L {\mt^2\over Q^2(\sqt,\mt)}\R \*\ln\L {\sqt\over
   Q^2(\sqt,\mt)}\R \nn\\
 &+&\ln\L {\mu^2\over \sqt}\R \*\ln\L {\mt^2\over \sqt}\R
  +{1 \over 2} \*\ln\L{\mu^2\over \sqt}\R ^2
  +{6\over 2}\*\ln\L {\mu^2\over \sqt }\R \nn\\
  &+&    \ln\L {\sqt\over Q^2(\sqt,\mt)}\R 
 - 2 \Li2{{\sqt\over Q^2(\sqt,\mt)}}
 - {\mt^2\over \sqt}\*\ln\L{\mt^2\over Q^2(\sqt,\mt)}\R 
 \nn\\&-&
 3\ln\L {Q(\sqt,\mt)-\mt\over
   Q(\sqt,\mt) }\R  -{3\mt\over Q(\sqt,\mt)+\mt} + {\pi^2\over 6}.
\end{eqnarray}
Following \Ref{Catani:1996vz} we used the bra-ket notation to represent the
leading-order amplitude as vector in colour space:
\begin{displaymath}
  \left| q\qb\to t\tb \,\ket.
\end{displaymath}
In the derivation we used the following result for 
$\Ioperator$ 
\begin{eqnarray}
  &&\hspace{-1cm}\Ioperator_q(\e,\mu^2;\{\},p_a) = 
  -{\as\over 2\pi} {(4\pi)^\e\over \Gamma(1-\e)}\Bigg\{
 {1\over \CF} T_t T_q \Bigg[\CF \L {\muq\over \sqt }\R ^\e
  \L \calV_Q(\sqt ,\mt,0; \e,\kappa)-{\pi^2\over 3}  \R \nn\\
  &+&
   \Gamma_Q(\mu,\mt;\e)+\gamma_q\*\ln\L {\muq\over \sqt }\R 
  +\gamma_q+K_q\Bigg]\nn\\
  &+& {1\over \CF} T_t T_q \Bigg[
  \CF\L {\muq\over \sqt }\R ^\e \L \calV_q(\sqt ,0,\mt;\e,2/3)-{\pi^2\over 3}\R
  +{\gamma_q\over \e} + \gamma_q\*\ln\L {\muq\over \sqt }\R  +
  \gamma_q+K_q\Bigg]
  \nn\\
  &+& (t \leftrightarrow \tb ) \Bigg\},
\end{eqnarray}
which can be easily obtained from \Ref{Catani:2002hc}.
The definitions of 
$\calV_Q,\calV_q, \Gamma_Q,\gamma_q,K_q$ can be found in
\Ref{Catani:2002hc}. 
The result for $\Ioperator_\qb$ can be obtained from the above result
by the replacement $(q \leftrightarrow \qb)$. 
The $T_t, T_q$ appearing in the above equation are colour-charge
operators, which act on the leading-order amplitudes that are vectors in 
colour space.
The calculation of this
specific contribution was further simplified by noting that, because of
the simple colour structure of the process at hand, the following
relation holds:
\begin{eqnarray}
  &&  \bra q\qb\to t\tb \left| T_q T_t \right| q\qb\to t\tb \,\ket
  = -\bra q\qb\to t\tb \left| T_q T_\tb\right| q\qb\to t\tb
  \,\ket
  \nn\\
  &&= -\bra q\qb\to t\tb\left|T_\qb T_t\right| q\qb\to t\tb \,\ket
  =\bra q\qb\to t\tb \left| T_\qb T_\tb \right| q\qb\to t\tb\,\ket.
  \label{eq:ColourHelp}
\end{eqnarray}
The square of the colour-correlated tree amplitudes is given by
\begin{eqnarray}
  &&\bra q\qb\to t\tb \left| T_q T_t   
  \right| q\qb\to t\tb \,\ket\nn\\
  &=& - 32\pi^2\alpha\as (N^2-1)\* {s\over s-\mz^2}
   \bigg(
   (d-2-\beta^2 (1-z^2))\*\gvq\*\gvt
   +\beta\*z\*(d-2)\*(d-3)\*\gaq\*\gat\bigg)\nn\\
   & = & 4\pi B,
   \label{eq:ColourCorrelatedLO}
\end{eqnarray}
with $B$ as defined in \Eq{eq:Bdef}. Comparing \Eq{eq:IRdivcontribution} with 
Eqs. (\ref{eq:Icontr}), (\ref{eq:ColourCorrelatedLO}), it is easy to
see that when combining the real and virtual corrections the IR
singularities indeed cancel. In addition, as promised already in
the  previous section, we see that indeed no $\e/\e$ terms
appear, which must be due to the `$d$-dimensional' factorization of 
the infrared singularities.
The $\Koperator$ and $\Poperator$ operators can be calculated along the same
lines as described above for the $\Ioperator$ operator. In particular 
we obtain 
\begin{eqnarray}
  \Koperator_q = -{\as\over 2\pi} T_t T_q \bigg\{ K(x,\sqt,\mt) -
  K(x,\sqtb,\mt) 
  \bigg\},
\end{eqnarray}
with 
\begin{eqnarray}
  && K(x,\sqt,\mt) = \bigg[J_{gQ}\left(x,{\mt^2\over\sqt }\right)\bigg]_+
  + 2 \bigg[{1\over 1-x}\bigg]_+ 
  \ln\left({(2-x)\sqt\over (2-x)\sqt+\mt^2}\right)\nn\\
  &+&
  \delta(1-x)\Bigg[{\mt^2\over \sqt}\ln\left(\mt^2\over \sqt
    +\mt^2\right) + {1\over 2} \*{\mt^2\over \sqt+\mt^2}
  + {3\over 2}  {2\mt\over \sqrt{\sqt +\mt^2}+\mt}\nn\\
  &+&      
  {3\over 2} 
  \ln\left({\sqt-2\mt\sqrt{\sqt+\mt^2}+2\mt^2 \over \sqt}\right)
  \Bigg]
  +
  P^{qq}_{\rm reg}(x)\ln\left({(1-x)\sqt\over
      (1-x)\sqt+\mt^2}\right),
  \nn\\
\end{eqnarray}
where the regular part $P^{qq}_{\rm reg}$ of the evolution kernel 
$P^{qq}(x)$ is given by
\begin{equation}
  P^{qq}_{\rm reg}(x) = - (1 + x),
\end{equation}
and the Plus prescription defines distributions in the usual way, through
\begin{equation}
\PP{F(z)} = \lim_{\eta\rightarrow 0}
  \left\{ \Theta(1-z-\eta) F(z) - \delta(1-z-\eta) \int_0^{1-\eta} F(y) dy
   \right\}.
\end{equation}
The function $\PP{J_{gQ}}$ is given in Eq.~(5.58) of
\Ref{Catani:2002hc}, 
and we reproduce it here explicitly:
\begin{eqnarray}
  \PP{J_{gQ}(x,y^2)} &=& \PP{{1-x\over 2 (1-x+y^2)^2} 
    - {2\over 1-x}\L 1+\ln(1-x+y^2)\R} \nn\\
  &+& \PP{{2\over 2-x}}\ln(2+y^2-x).
\end{eqnarray}
For the $\Poperator$ operator we find
\begin{eqnarray}
  \Poperator_q =  {\as\over 2\pi} P^{qq}(x) T_tT_q
   \ln\L {\sqtb\over  \sqt}\R,
\end{eqnarray}
with 
\begin{equation}
  P^{qq}(x) = P^{qq}_{\rm reg}(x) + \PP{2\over 1-x} + {3\over 2}\delta(1-x).
\end{equation}
Note that in deriving the above relations 
we used again relation \Eq{eq:ColourHelp} to simplify the colour charge
algebra. The corresponding results for the antiquark give the same
contribution for the $\Koperator$ and $\Poperator$ operators. 
Note that when calculating
\begin{equation}
 \int dx \int_2 d\sigma_{\rm LO}(x\kq,\kqb,\kt,\ktb)\otimes (
 \Koperator_q
+ \Poperator_q ),
\end{equation}
one has to replace $\kq$ by $x\kq$ in the calculation of the 
colour-correlated matrix element \Eq{eq:ColourCorrelatedLO} as well as
in the phase-space measure. For details concerning the evaluation of
the Plus prescriptions we refer to \Ref{Catani:2002hc}.
A remark might be in order concerning the appearance of the
$\Koperator$ and $\Poperator$ operators. To the order we are working
here, there is no contribution from the factorization of initial-state
singularities. At first sight the fact that the evolution kernel $P^{qq}$
appears might thus look a bit strange. The reason is just that a
corresponding term is included in the dipole contribution, which is
combined with the real corrections. If we apply the 
subtraction method as it is described in \Ref{Catani:2002hc}, we thus have to
consider the contributions from the $\Koperator$ and $\Poperator$
operators
as shown above.
In principle one could think of changing slightly the form of 
the subtraction terms; but in
that case the analytic integration over the subtraction terms would also
need to be redone. Let us close this
section with some remarks about the subtraction term 
\begin{equation}
  \L d\sigma_{\rm LO} \otimes 
    dV_{\rm dipole}\R (\kq,\kqb,\kt,\ktb,\kg)
\end{equation}
in \Eq{eq:Master}, which is combined with the real corrections.
This contribution is obtained as a sum over individual `dipoles'
${\cal D}^a_{ij}$, ${\cal D}^{ij}_a$:
\begin{eqnarray}
  \L d\sigma_{\rm LO} \otimes 
    dV_{\rm dipole}\R 
    &=& {1\over 2 s} {1\over 4 N^2}\int dR_3(\kt,\ktb,\kg) 
    \bigg( {\cal D}^q_{tg} + {\cal D}^\qb_{tg} + {\cal D}^q_{\tb g} 
    + {\cal D}^\qb_{\tb g}\nn\\
    &+& {\cal D}^{qg}_t + {\cal D}^{\qb g}_t + {\cal D}^{qg}_\tb
    + {\cal D}^{\qb g}_\tb\bigg).
\end{eqnarray}
Here $i$ and $j$ are the unresolved partons, while $a$ plays the
r\^ole of a spectator.
The explicit expressions for the 8 dipoles can be easily obtained from 
\Ref{Catani:2002hc}. For example, we get 
\begin{eqnarray}
  {\cal D}^q_{tg} &=& - {1\over (\kt+\kg)^2 - \mt^2}
  {1\over x_{tg,q}} V_{tg}^q \nn\\
  &\times&\bra q(\tilde \kq)\qb(\kqb) \to t(\tilde\kt)\tb(\kt) \bigg| 
  T_q\cdot T_t
   \bigg| q(\tilde \kq)\qb(\kqb) \to t(\tilde \kt)\tb(\kt)\ket\nn\\
   &=&- {1\over (\kt+\kg)^2 - \mt^2}
  {1\over x_{tg,q}} V_{tg}^q \,\,B\bigg|_{\kq = \tilde \kq, \kt=\tilde \kt },
\end{eqnarray}
where
\begin{equation}
  x_{tg,q} = {\kq\cdot\kt +\kq\cdot\kg - \kt\cdot\kg
    \over \kq\cdot\kt +\kq\cdot\kg}, 
\end{equation}
\begin{equation}
  \tilde \kq^\mu = x_{tg,q}\kq^\mu, \quad
  \tilde \kt^\mu = \kt^\mu+\kg^\mu - (1-x_{tg,q})\kq^\mu ,
\end{equation}
and
\begin{equation}
  V_{tg}^q = 8\pi\as\bigg\{
  {2\over 2-x_{tg,q}-\tilde z_t} - 1 - \tilde z_t - {\mt^2\over \kt\cdot\kg}
  \bigg\}.
\end{equation}
The four-momentum of parton $i$ is denoted by $k_i$.
The momentum fraction $\tilde z_t$ is defined  by
\begin{equation}
  \tilde z_t={\kq\cdot\kt 
    \over \kq\cdot\kt +\kq\cdot\kg}.
\end{equation}
For the remaining dipoles the  obtained results are similar.

The numerical implementation of the subtractions terms shown above  is 
straightforward. So far we have only discussed the calculation of the
total cross section. In principle also more exclusive quantities can
be calculated without any significant change. In the 
next section we will discuss our
numerical results for the cross section. More exclusive quantities
will be discussed elsewhere.

\section{Numerical results}
\label{sec:results}
\begin{figure}[!htbp]
  \begin{center}
    \leavevmode
    \parbox{\textwidth}{\large\centerline{%
        $\raisebox{-1.25cm}{\includegraphics{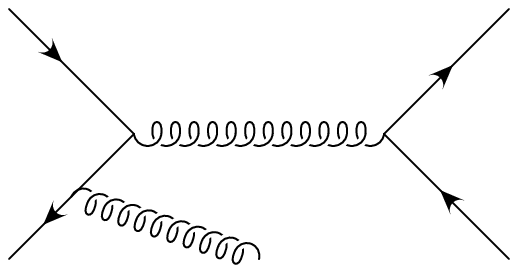}}
        \,\times
        \,\left(\,\,\raisebox{-1.25cm}{\includegraphics{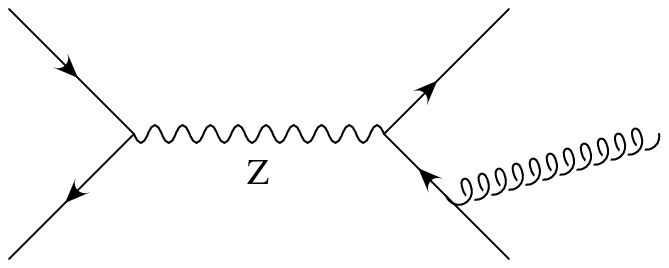}}
        \right)^{\mbox{\Large$\ast$}}$}
      \normalsize\vspace*{0.2cm}

      \centerline{a)}
      }
    \vspace*{0.3cm}

    \parbox{\textwidth}{\large\centerline{%
        $\raisebox{-1.25cm}{\includegraphics{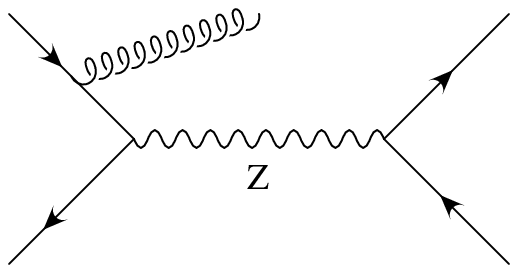}}
        \,\times
        \,\left(\,\,\raisebox{-1.25cm}{\includegraphics{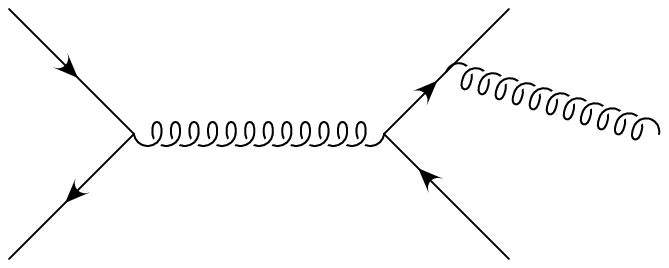}}
        \right)^{\mbox{\Large$\ast$}}$}
      \vspace*{0.2cm}

       \normalsize\centerline{b)}
      }
    \caption{Schematic representation of real corrections to the
        QCD-box contribution (a) 
        and the EW-box contribution (b).}
    \label{fig:EW-QCD-splitting}
  \end{center}
\end{figure}
In the following we will discuss numerical results for the weak
corrections to the total cross
section. If not stated otherwise, we used the following
values for the masses 
\begin{displaymath}
  \mz = 91.1876 \GeV,\quad
  \mw = 80.425  \GeV,\quad
  \mh = 120\GeV,\, 
\end{displaymath}
\begin{displaymath}
  \mb = 4.82\GeV,\quad
  \mt = 178.0 \GeV,  
\end{displaymath}
and
\begin{displaymath}
  \alpha(2\mt) = {1\over 126.3},\quad
  \as = 0.1,\quad
  \sw^2 = 0.231
\end{displaymath}
for the couplings. Using $\mz$ and $\mw$ as input parameters, the weak
mixing angle can be in principle calculated within the
theory. However, since our calculation is leading order in the
electroweak coupling, we expect that the numerical choice for $\sw$ as
given within the $\overline{\mbox{MS}}$-scheme, will give results
closer to the actual values. 
\begin{figure}[!htbp]
  \begin{center}
    \leavevmode
    \includegraphics[width=12cm]{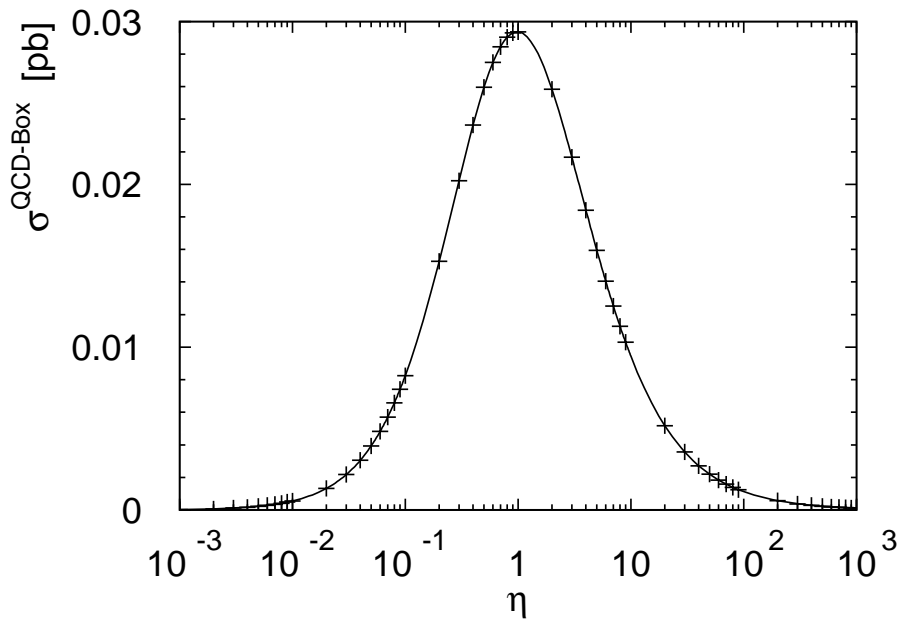}
    \caption{Comparison with results available in the literature 
      \cite{Kniehl:1989qu}.}
    \label{fig:LiteraturCompare}
  \end{center}
\end{figure}
Before showing our final results for the cross section, let us first 
discuss several checks we performed.
The analytic expressions for the vertex corrections to the initial vertex 
can be found in the literature. We compared 
with the results given in \Refs{Grzadkowski:1986pm,Bohm:1986rj}
and found complete agreement. For the corrections to the
final vertex, no such compact expressions can be found in the
literature. Using the same
input parameters as in \Ref{Beenakker:1993yr} we compared  with plots shown in
\Ref{Beenakker:1993yr} and
found agreement. Furthermore 
a precise numerical comparison with Bernreuther, Fücker and Si 
\cite{BeFuSi} who also
recently finished an independent calculation of the
weak corrections, lead to complete agreement. For the case of  
the box diagrams it is possible to compare the contribution of the
QCD boxes with an analytic result available in the literature 
\cite{Kniehl:1989qu}.
\begin{figure}[!htbp]
  \begin{center}
    \leavevmode
    \includegraphics[width=12cm]{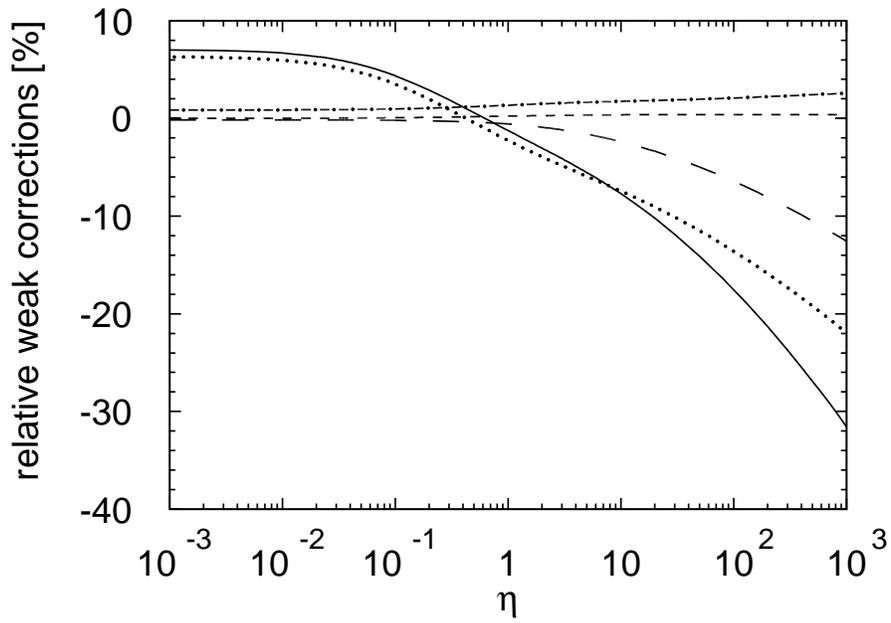}
    \caption{Different contributions to the electroweak corrections
      for incoming up-quarks: 
      Initial vertices (long-dashed), final vertices (dotted), 
      EW-box (dash-dotted), QCD-box (dashed). The sum is shown as a full line.}
    \label{fig:Contributions}
  \end{center}
\end{figure}
In \Ref{Kniehl:1989qu} the corrections to the total cross section 
were calculated
using the optical theorem. In our calculation we just need to split
the real corrections into the  contribution belonging to the EW-box
and that belonging to the QCD box. 
As far as the matrix elements
for the real corrections are concerned, the contribution where the
momentum flow in the $Z$-propagator is the total centre-of-mass energy
belongs to the QCD box. On the other hand, the contribution 
where the momentum of the gluon propagator is equal to the total
centre-of-mass energy belongs to the EW-box contribution. 
Sample diagrams for both contributions are shown 
Fig.~\ref{fig:EW-QCD-splitting}.
For the
subtraction terms there is no such difference, they have to be
distributed equally. This splitting might sound somewhat
artificial, but it allows a direct comparison with \Ref{Kniehl:1989qu}. In
Fig.~\ref{fig:LiteraturCompare} we show the analytic result from 
\Ref{Kniehl:1989qu} as a line. The
crosses are obtained from the numerical integration of our results
for the QCD box over the full phase space. We find
complete agreement taking the numerical uncertainties of the
phase-space integration into account. This is a highly non-trivial test,
because the entire contribution from the subtraction method is checked.
In addition we compared again with the results by
Bernreuther, Fücker and Si and found perfect agreement \cite{BeFuSi}.  

Note that for the box contributions only the axial-vector part
(proportional to $\gat\gaq$) contributes
to the total cross section thanks to the Furry theorem. We included the
vector part  (proportional to $\gvt\gvq$)
in our calculation as well, because our aim is to allow also the 
calculation of differential quantities (with or without cuts) where these 
terms might contribute. We checked that for the total cross section
the vector part indeed cancels in the numerical evaluation of the
phase-space integrals --- providing a further check of our numerical 
implementation. 
Terms proportional to $\gvt\gaq$ or $\gat\gvq$ contribute to parity
violating
observables only. They are relevant for spin dependent quantities and
have not been included in the present analyses.
An important consequence of the Furry theorem
is that  the result for incoming
down-type quarks can be obtained directly from the one for up-type quarks as
far as the boxes are concerned. There is just a relative sign between the
two contributions, because of the sign difference in the weak
isospin. 
At the hadron level
the contribution of the box diagrams is thus directly proportional to the
difference of the parton distribution functions between up- and
down-type quarks. This leads to a suppression of the contribution of the 
box diagrams.

Let us now discuss the numerical results for the cross section.
In Fig.~\ref{fig:Contributions} we show the separate contributions as
well as the sum for the partonic cross section for incoming up-quarks
as a function of 
\begin{equation}
  \eta = {s\over 4\mt^2} - 1.
\end{equation} 
For a Higgs mass of $\mh=120\GeV$ used in 
Fig.~\ref{fig:Contributions} the  dominant contribution is given by 
the vertex corrections.
It can also be seen that the contribution from EW-boxes is much larger
than the contribution from the QCD-boxes. 
We have checked that the purely weak contributions, of order
$\alpha^2$ are completely negligible. At hadron colliders the parton 
subenergies may reach the order of TeV and beyond. In this region the
suppression of the cross section by large Sudakov logarithms starts to
become important, similiarly to the situation for purely electroweak 
processes (see e.g. \Refs{Kuhn:1999nn,Kuhn:2001hz}). In this region the weak corrections
are of the order of 10\% and more. 
\begin{figure}[!htbp]
  \begin{center}
    \leavevmode
     \includegraphics[width=12cm]{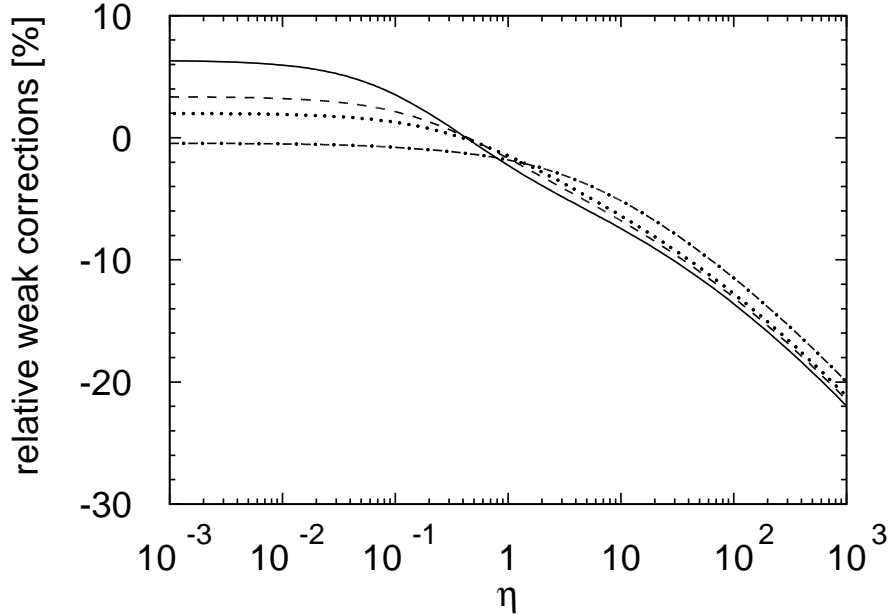}
    \caption{Relative change of the cross section from corrections to
      the final vertex for different Higgs masses
      $\mh=120\GeV$ (full line),
      $\mh=180\GeV$ (dashed), $\mh=240\GeV$ (dotted), 
      $\mh=1000\GeV$ (dashed-dotted). }
    \label{fig:MH-dependence}
  \end{center}
\end{figure}
Furthermore, the vertex corrections
depend strongly on the Higgs mass as shown in
Fig.~\ref{fig:MH-dependence}, and this
dependence is particularly pronounced in the threshold region.
(For a related discussion see \Ref{Jezabek:1993eh}.) In particular for
small $\mh$ and small velocity $\beta$ one is still sensitive to the 
attractive Yukawa force.
For large $\mh$ and small $\beta$, on the other hand, the Higgs
contribution vanishes. 
\begin{figure}[!htbp]
  \begin{center}
    \leavevmode
    \includegraphics[width=12cm]{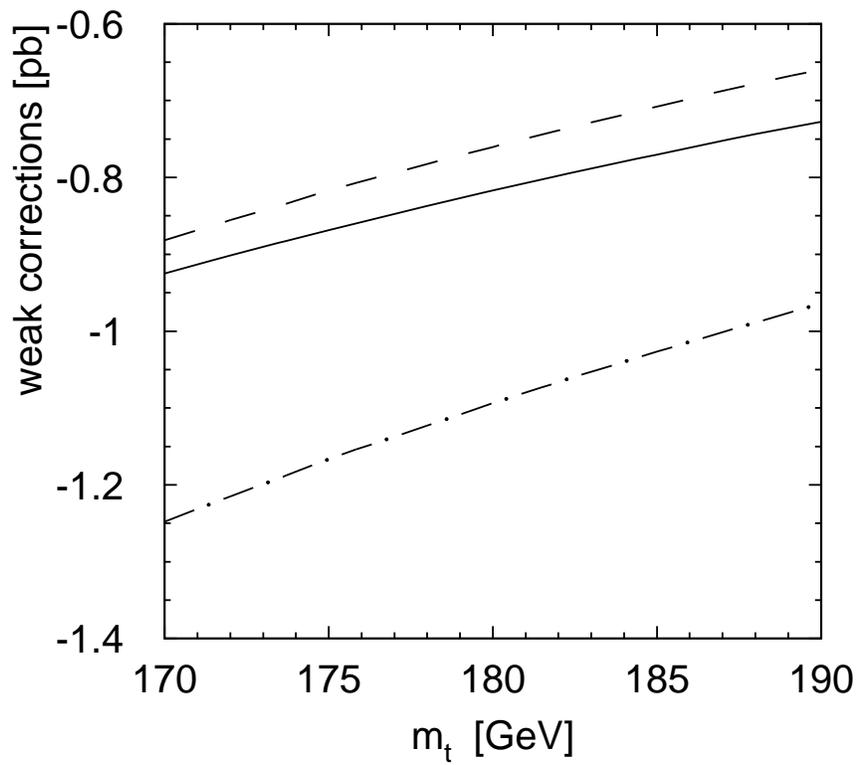}
    \caption{Dependence of the $q\bar q$ induced hadronic cross 
      section on the top mass $\mt$ (LHC) for 3 different Higgs masses
      ($\mh=120$ solid line, $\mh=200$ dashed line, $\mh=1000$
      dashed-dotted line).}
    \label{fig:TopmassDependenceLHC}
  \end{center}
\end{figure}
\begin{figure}[!htbp]
  \begin{center}
    \leavevmode
    \includegraphics[width=12cm]{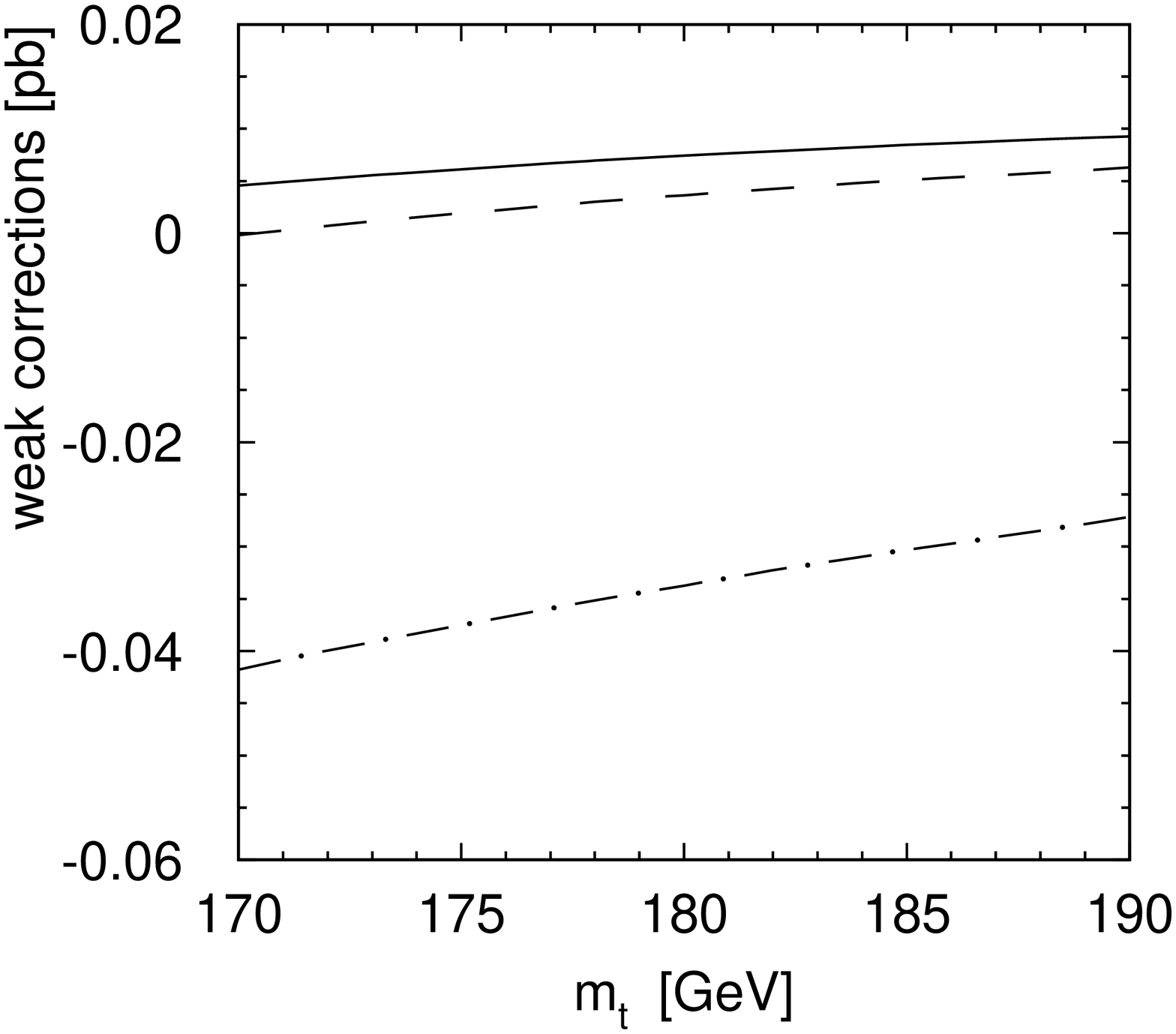}
    \caption{Dependence of the $q\bar q$ induced hadronic cross 
      section on the top mass $\mt$ (Tevatron) for 3 different Higgs
      masses ($\mh=120$ solid line, $\mh=200$ dashed line, $\mh=1000$
      dashed-dotted line).}
    \label{fig:TopmassDependenceTEV}
  \end{center}
\end{figure}
In Figs.~\ref{fig:TopmassDependenceLHC} and
\ref{fig:TopmassDependenceTEV} 
the total contribution of the weak corrections to
the quark--antiquark induced part of the hadronic cross section is shown,
which should be compared to the QCD corrected total cross section of 
5.75~pb \cite{Cacciari:2003fi} and  
833~pb \cite{Bonciani:1998vc} for 1.96 TeV and 14 TeV respectivly. As far
as the total cross section is concerned, the effects are evidently very small.
Electroweak corrections are, however, important for differential distributions,
which are enhanced in the region of large parton subenergies \cite{KSU}.

\section{Conclusion}
\label{sec:concl}
In this article we have evaluated the complete electroweak corrections to
top-quark pair production in quark--antiquark annihilation including
terms from the interference between QCD and electroweak amplitudes. In
particular we present short analytical results, which will be 
useful for further
investigations. As a first application we have studied the impact of
the weak corrections on the total cross section. We confirm the
findings of \Ref{Beenakker:1993yr} that the correction to the total cross
section is negligible.

{\bf Acknowledgments:} 
We would like to thank W.Bernreuther, M. Fücker and Z.-G. Si for
useful discussions and for a detailed comparison of results prior 
to publication. 

\appendix

\section{List  of used integrals}
\label{sec:integrals}
%
%
Using the definitions
\begin{eqnarray*}
  &&B_0(p_1^2,\m1^2,\m2^2)={1\over i\pi^2} \int d^d\ell 
  {(2\pi\mu)^{2\e}\over (\ell^2-\m1^2+i\e)
    ((\ell+p_1)^2-\m2^2+i\e)}
  \nn\\
  &&C_0(p_1^2,p_2^2,p_1\cdot p_2,
  \m1^2,\m2^2,\m3^2)=\nn\\
  &&{1\over i\pi^2} \int d^d\ell 
  {(2\pi\mu)^{2\e}\over (\ell^2-\m1^2+i\e)
    ((\ell+p_1)^2-\m2^2+i\e)((\ell+p_1+p_2)^2-\m3^2+i\e)}
\end{eqnarray*}
the integrals used in section \ref{sec:virtual} are
\begin{eqnarray}
  B_0^1(1,3)  &=& B_0(s,0,\mz^2) \\
  B_0^1(1,4) &=& B_0(\mt^2,0,\mt^2) \\
  B_0^1(3,4)  &=& B_0(\mt^2,\mz^2,\mt^2) \\
  B_0^1(2,4) &=& B_0(-{s\over 2}\*(1-\beta\*z)+\mt^2,0,\mt^2) \\
  B_0^2(2,4) &=& B_0(-{s\over 2}\*(1+\beta\*z)+\mt^2,0,\mt^2) \\
  B_0^3(1,3) &=& B_0(s,\mt^2,\mt^2)\\
  B_0^4(1,2) &=& B_0(\mt^2,\mb^2,\mw^2)\\
  B_0^4(1,3) &=& B_0(s,\mb^2,\mb^2)\\
  B_0^5(1,2) &=& B_0(\mt^2,\mt^2,\mh^2)\\
  B_0^6(1,3)  &=& B_0(s,0,0) \\
  C_0(1,3,4) &=& C_0(s,\mt^2,-{s\over 2},0,\mz^2,\mt^2)\\
  C_0^1(1,2,4) &=& C_0(0,-{s\over 2}\*(1-\beta\*z)+\mt^2,{s\over 4}\*(1-\beta\*z),0,0,\mt^2)\\
  C_0^2(1,2,4) &=& C_0(0,-{s\over 2}\*(1+\beta\*z)+\mt^2,{s\over 4}\*(1+\beta\*z),0,0,\mt^2)\\
  C_0^1(2,3,4) &=& C_0(0,\mt^2,-{s\over 4}\*(1-\beta\*z),0,\mz^2,\mt^2)\\
  C_0^2(2,3,4) &=& C_0(0,\mt^2,-{s\over 4}\*(1+\beta\*z),0,\mz^2,\mt^2)\\
  C_0^{\rm QCD1}(1,3,4) &=& C_0(s,\mt^2,-{s\over 2},0,0,\mt^2)\\
  C_0^3 &=& C_0(\mt^2,\mt^2,{s\over 2}-\mt^2,\mt^2,\mz^2,\mt^2) \\
  C_0^4 &=& C_0(\mt^2,\mt^2,{s\over 2}-\mt^2,\mb^2,\mw^2,\mb^2) \\
  C_0^5 &=& C_0(\mt^2,\mt^2,{s\over 2}-\mt^2,\mt^2,\mh^2,\mt^2) \\
  \DEWsix1 &=& 
  D_0^6(0,0,\mt^2,{s\over2},-{s\over4}\*(1-\beta\*z),-{s\over4}\*(1+\beta\*z),0,0,\mz^2,\mt^2)\\
   \DEWsix2 &=& 
  D_0^6(0,0,\mt^2,{s\over2},-{s\over4}\*(1+\beta\*z),-{s\over4}\*(1-\beta\*z),0,0,\mz^2,\mt^2)\\
  \Dsix1 &=& 
  D_0^6(0,0,\mt^2,{s\over2},-{s\over4}\*(1-\beta\*z),-{s\over4}\*(1+\beta\*z),0,0,0,\mt^2)\\
  \Dsix2 &=& 
  D_0^6(0,0,\mt^2,{s\over2},-{s\over4}\*(1+\beta\*z),-{s\over4}\*(1-\beta\*z),0,0,0,\mt^2)
 \end{eqnarray}


\newcommand{\zp}{Z. Phys. }\def\as{\alpha_s }\newcommand{\prd}{Phys. Rev.
  }\newcommand{\pr}{Phys. Rev. }\newcommand{\prl}{Phys. Rev. Lett.
  }\newcommand{\npb}{Nucl. Phys. }\newcommand{\psnp}{Nucl. Phys. B (Proc.
  Suppl.) }\newcommand{\pl}{Phys. Lett. }\newcommand{\ap}{Ann. Phys.
  }\newcommand{\cmp}{Commun. Math. Phys. }\newcommand{\prep}{Phys. Rep.
  }\newcommand{\jmp}{J. Math. Phys. }\newcommand{\rmp}{Rev. Mod. Phys. }

\end{document}